\documentclass[11pt]{article}
\usepackage{epsfig}

\def\Title#1{\begin{center} {\Large {\bf #1} } \end{center}}
\long\def\AUTHORS#1{ #1\\[3mm]}
\long\def\AFFILIATION#1#2{$^{#1}\,$ #2\\}
\begin{document}
%{\small \it APS DPF 2017, Fermi Lab}
\vspace{12pt}
\def\Title#1{\begin{center} {\Large {\bf #1} } \end{center}}

%\begin{document}
\Title{Survey of the physics landscape and attempts to improve diversity$^2$}

\bigskip\bigskip

%+\addtocontents{toc}{{\it D. Reggiano}}
%+\label{ReggianoStart}
\AUTHORS{\underline{B. Beckford}$^1$ }
%%%
{\small \it 
\AFFILIATION {1}{Department of Physics, University of Michigan, Ann Arbor, MI 48109 USA}}
\AFFILIATION {2}{Talk presented at the APS Division of Particles and Fields Meeting (DPF 2017), July 31-August 4, 2017, Fermilab. C170731}

% Enter contact e-mail address here.
\centerline{Contact email: {\it bobeck@umich.edu}}
%\begin{raggedright}  
%{\it Brian Beckford\index{}\\
%Department of Physics\\
%University of Michgan\\
%Ann Arbor, MI, 48109}
%\bigskip\bigskip
%\end{raggedright}

\begin{center}
\section*{Abstract}
\end{center}
Students from statistically underrepresented minority (URM) groups and women earn a smaller fraction of undergraduate and graduate degrees in most physical sciences, particularly physics. This underrepresentation is also prevalent at the faculty level and in higher administration roles at most physics departments, universities, and national laboratories. This proceedings summarizes a presentation that presented statistics on participation in physical sciences, and discussed the outcomes of initiatives such as Bridge Programs which aim to improve diversity in physics graduate programs.\\

\section{Introduction}
In last few years remarkable advances and discoveries in physics and astronomy have been made ranging from small to large scales with the discovery of the Standard Model Higgs boson to the detection of gravitational waves by LIGO~\cite{Higgs,LIGO}. At such a time of continued advances in technology coupled with an increasing demand for a strong national STEM workforce, and an expanding U.S. minority population, physics as a field and community still lingers close the bottom of the physical sciences with regards to degrees earned by URM college age population. Of the current U.S. population women compose roughly 50\%, with Hispanic-Americans, Black/African-Americans, Asian-Americans, and (American Indian, Alaskan Native, Native Hawaiian, Pacific Islander, and those who reported more than one race making up 17\%, 13\%, 6\% and 2\% of the population respectively. However physics has held on tightly to the title of the least diverse of all physical sciences with respect to gender and underrepresented minorities (URM). 

National organizations including the American Physical Society (APS), the American Institute of Physics, and the American Association of Physics Teachers (AAPT) have recognized their role in addressing diversity and inclusion in the physics community. The APS Department of Education and Diversity created the APS Bridge Program, a national effort to increase the number of physics PhDs awarded to underrepresented minority (URM) students, defined by the project as Black/African-Americans, Hispanic-Americans and Native-Americans. Alongside this program the department has launched the National Mentoring Community, an effort to increase the number of underrepresented minority students who complete bachelor's degrees in physics and continues to expand the Conference for Undergraduate Women in Physics (CUWiP). APS has most recently started a forum on Diversity and Inclusion. In this report I will summarize current representation of groups in physics to further illustrate the landscape and discuss some outcomes to addressing diversity that bridge programs have achieved. 

\section{Motivation}
  \begin{figure}[htb]
	\begin{center}
	\epsfig{file=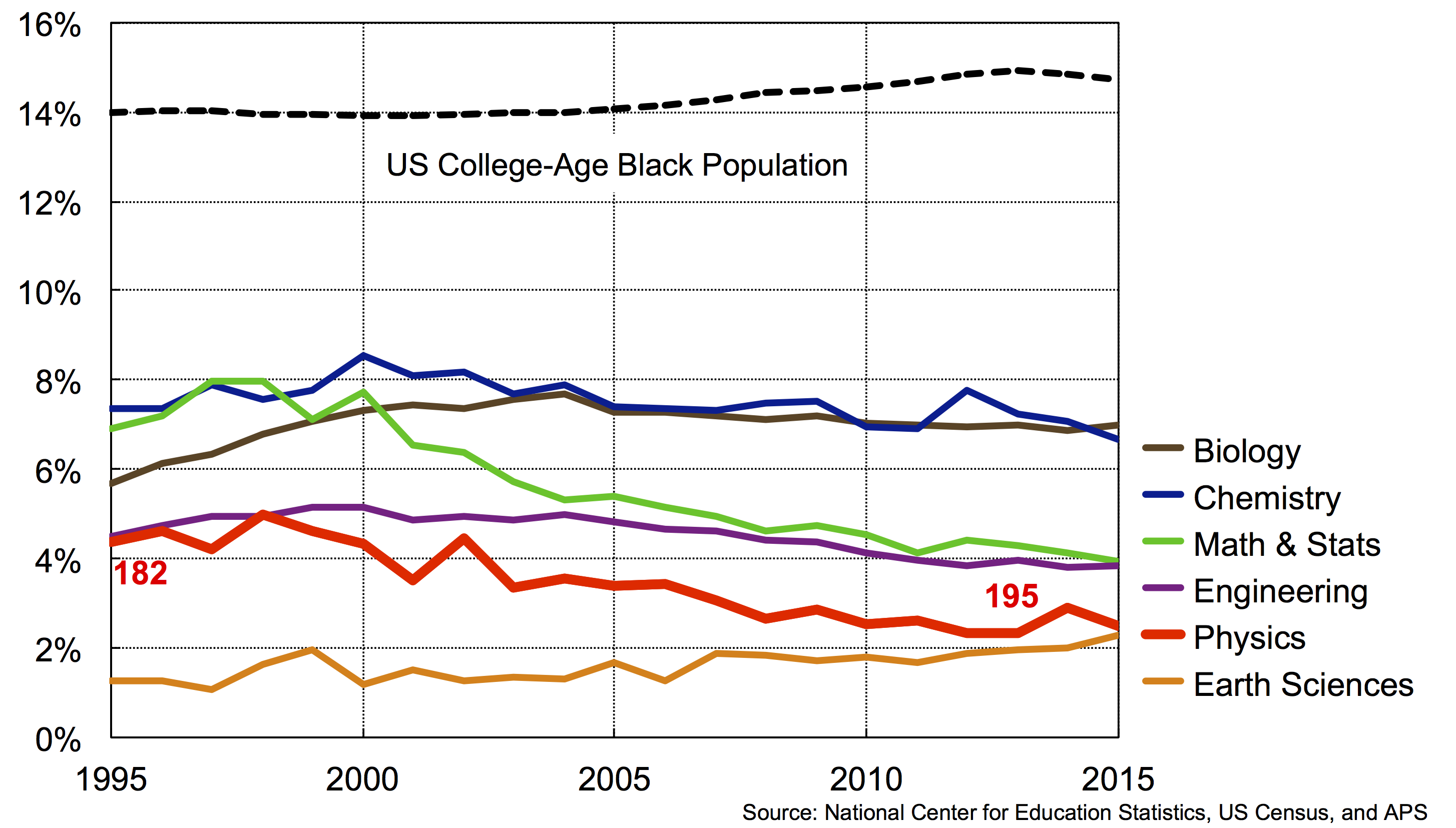,width=.8\columnwidth}
	\epsfig{file=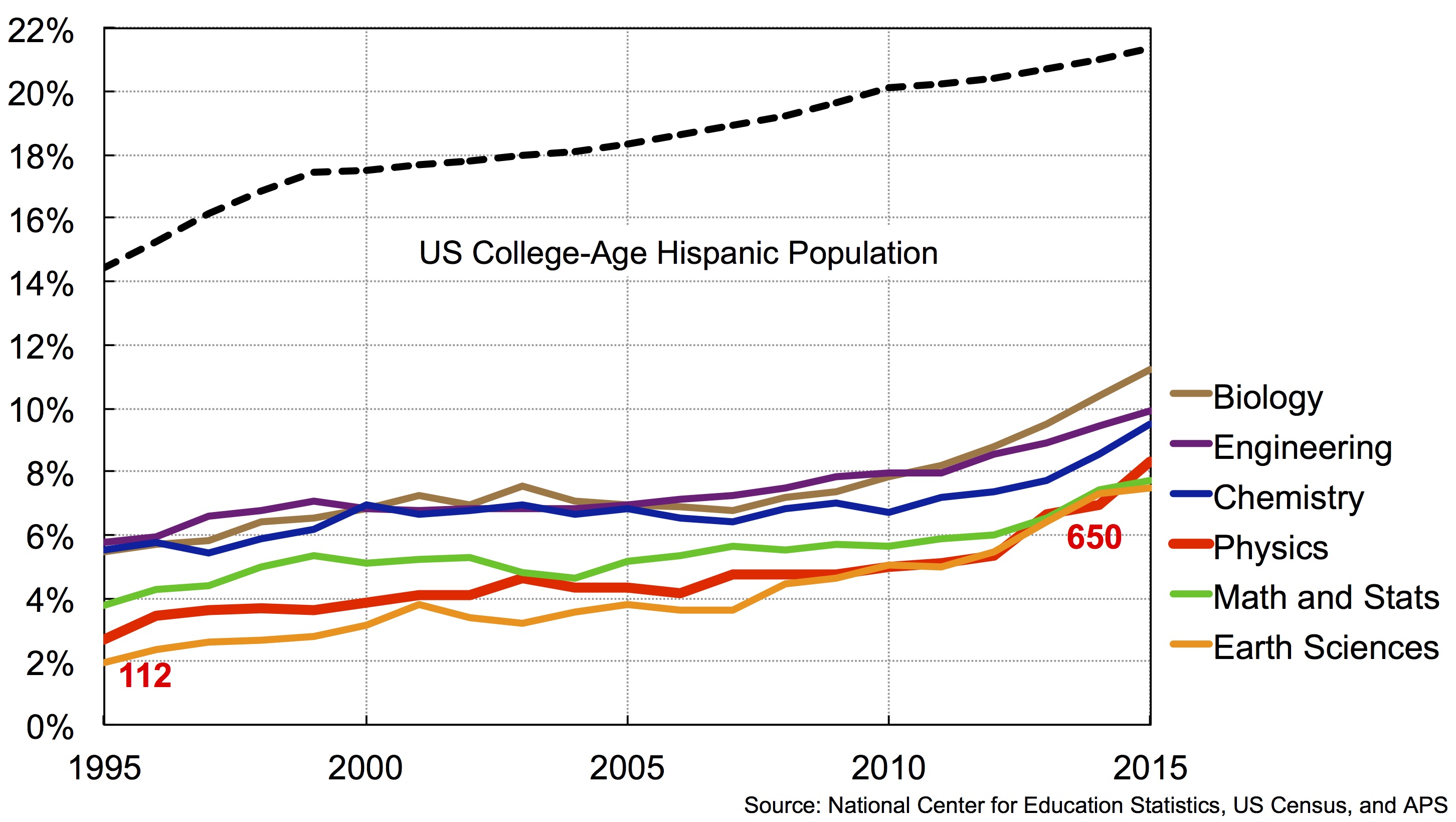,width=.8\columnwidth}
			\caption{(Color online) Percentage of bachelor's degrees in biology, chemistry, math $\&$ stats, engineering, physics, and earth sciences earned by Black/African-American and Hispanic Americans~\cite{AIP,APS}.}
  	\vspace{-.2cm}
	\label{fig:urm_physics_BS_AA_HA}
	\end{center}
\end{figure}

The percentage of bachelor's degrees from the US college-age populace earned by URMs in physics is close to the lowest fraction with only earth sciences accounting for less. A similar trend is evident for Hispanic American students as well. In Figure~\ref{fig:urm_physics_BS_AA_HA} the percentage of bachelor's degrees earned Blacks/African Americans and Hispanic Americans for the last 20 years is shown left to right respectively. The trends are presented for biology, chemistry, math $\&$ stats, engineering, physics, and earth sciences. URMs composed of Hispanic-, Black/African-, and Native- Americans account for 11\%, 10\% , and 6\% of BS, MS, PhDs degrees earned in physics~\cite{AIP,APS}. The fraction of BS degrees earned by Hispanic Americans has grown overall by 6\% along with the population. For Black/African American students the problem is dire as the population has remained relatively flat with a decrease of 2\% in the fraction of degrees earned.  Details about the participation of Native Americans in physical sciences can be referenced in the AIP report on Native American Participation among Bachelors in Physical Sciences and Engineering\cite{AIP}.

\begin{figure}[h!]
	\begin{center}
			\epsfig{file=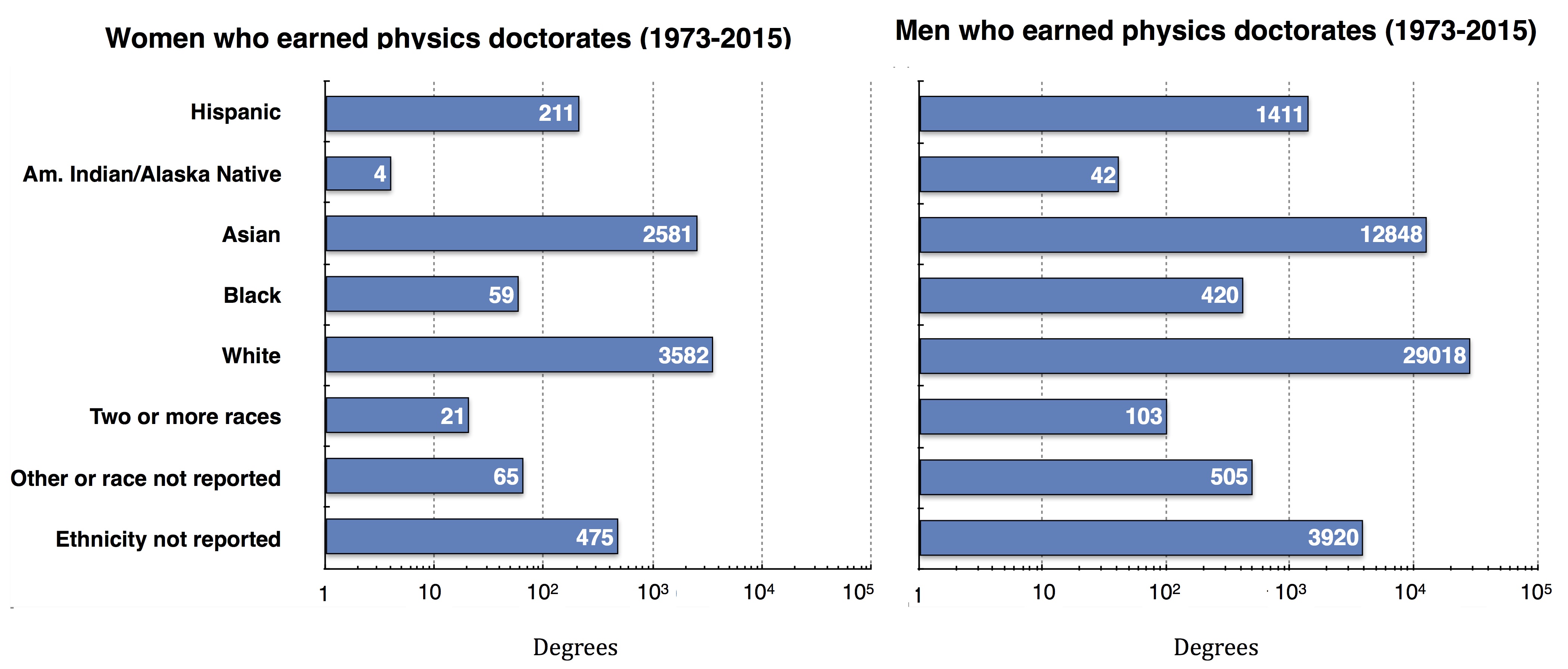,width=1.0\columnwidth}
			\caption{(Color online) The number of doctorate degrees earned from 1973-2015 for different demographics.}
  	\vspace{-.2cm}
	\label{fig:urm_degrees_earned_physics}
	\end{center}
\end{figure}

The number of doctorate degrees earned from 1973-2015 of different demographic groups, for men and women, is given in the left and right panels respectively in Figure~\ref{fig:urm_degrees_earned_physics}. The number degrees earned by white men and women since 1973 is seen to be clearly orders of magnitudes above those earned by Black./African Americans, Hispanic Americans, and Native Americans. Although Asian Americans are considered a racial/ethnic minority group in the U.S. they are not underrepresented in physics.

%  \begin{figure}[htb]
%	\begin{center}
%			\includegraphics*[width=.5\columnwidth]{NA_BS_degrees.PDF}
			%\includegraphics*[width=.45\columnwidth]{NA_BS_degrees.PDF}
%			\caption{(Color online) Percentage of bachelor's degrees in physical sciences earned by Native Americans~\cite{AIP,APS}.}
  %	\vspace{-.2cm}
%		\label{fig:urm_physics_BS_NA}
%	\end{center}
%\end{figure}

\section{Bridge Programs}
  \begin{figure}[htb]
	\begin{center}
			\caption{(Color online) Percentage of bachelor's and doctoral degrees earned by URM in physics. The APS Bridge Program aims to close the gap between the percentage of physics PhDs awarded to underrepresented minority students to match the percentage of physics Bachelor's degrees. The number of PhDs degree needed to close this achievement gap is roughly 30 more.}
  	\vspace{-.2cm}
	\label{fig:urm_physics}
	\end{center}
\end{figure}

Bridge Programs are an revolutionary approach to addressing the underrepresentation of some groups in physics. They aim to provide opportunities for students to be successful that may not have had such chances by traditional means. The APS Bridge Program has the goal to increase, within a decade, the percentage of physics PhDs awarded to underrepresented minority students to match the percentage of physics Bachelor's degrees granted to these groups as shown in Figure~\ref{fig:urm_physics}. It also aims to develop, evaluate, and document sustainable model bridging experiences that improve the access to and culture of graduate education for all students, with emphasis on those underrepresented in doctoral programs in physics~\cite{APS_BP}.  Currently, there are a few established Bridge Programs such as the Fisk-Vanderbilt bridge program, which was started around 2004, the Imes-Moore Bridge Program in Applied Physics at University of Michigan, Columbia university's bridge program, and the program at MIT.

\begin{figure}[ht]
	\begin{center}
		\epsfig{file=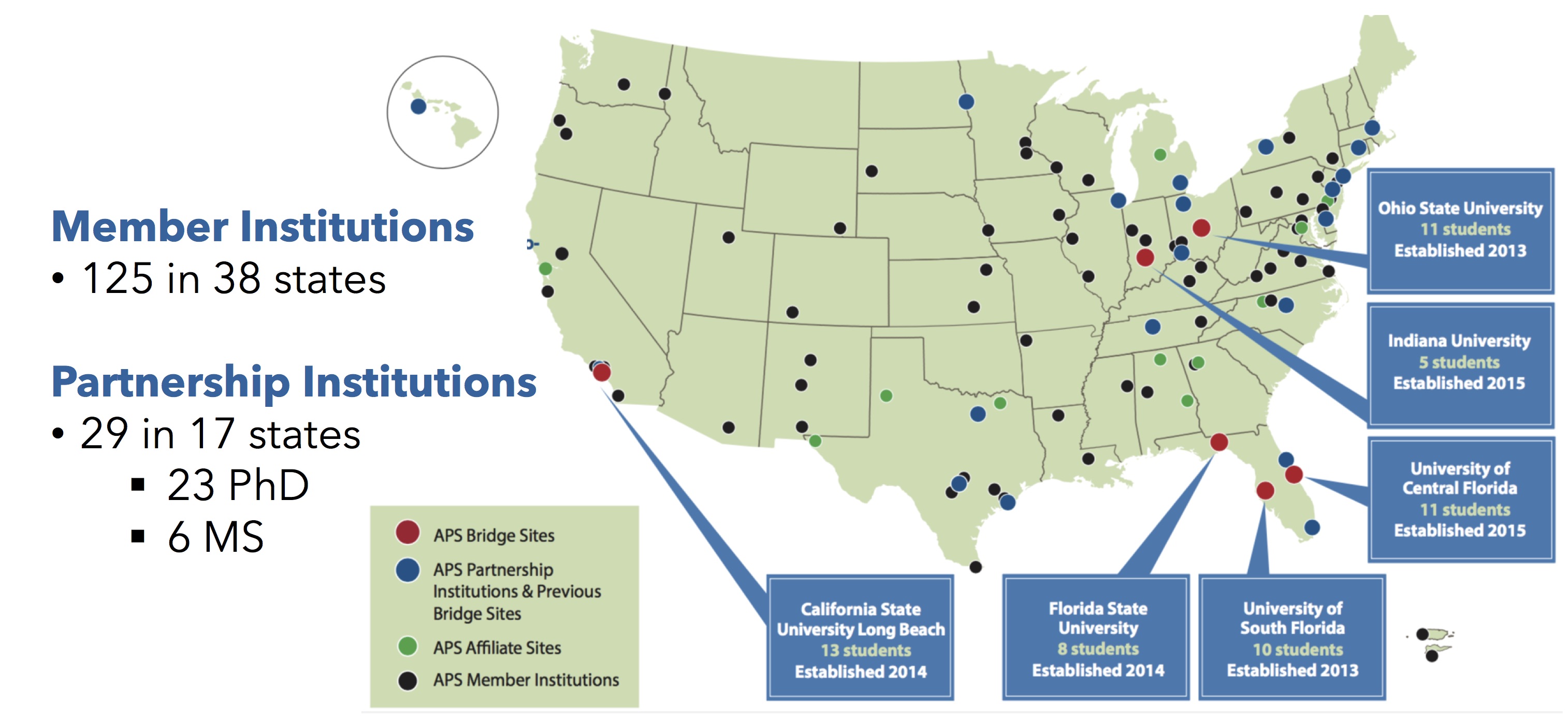,width=1.0\columnwidth}
			\caption{(Color online) National network of institutions participating in and affiliated with the APS Bridge Programs~\cite{APS_Connections}. The six APS funded bridge sites established in Ohio, Florida, Indiana, and California are indicated by the blue call-out boxes. }
  	\vspace{-.2cm}
	\label{fig:BP_Sites}
	\end{center}
\end{figure}

With the goal in mind of increasing the number of PhDs earned by underrepresented minority students the APS Bridge Program selected doctoral and master's granting institutions to be funded and establish sustainable programs to prepare students for entering and successfully completing graduate studies in physics. The model most generally adopted by Bridge Programs is to accept students into MS to PhD bridging experience. During the 1-2 year period the Bridge Programs support student in enhancing their academic and research  skills for acceptance into PhD programs. The six bridge sites in Indiana, Ohio, Florida, and California provide advanced coursework, research experiences, and substantial mentoring for  accepted students. Physics departments that are part of the nationwide coalition involved with the APS Bridge Program can be seen in Figure~\ref{fig:BP_Sites}, where the APS funded sites are highlighted. Interest in the APS Bride Program continues to swell, a result of   dissemination of the best strategies and program achievements. At present the APS Bridge Program has 125 Member Institutions in 38 states and 29 Partnership Institutions in 17 different states. Of the 29 partnership institutions, 23 are universities that grant a PhD in physics as their highest degree~\cite{APS_BP}. 

%\subsection{Key Components}
\subsection{Principle Components} Bridge and Partnership Institutions that have achieved success and seen in increase in URM participation seen in their departments have attributed it to numerous important components which have formed a list of best practices, some of which include: (1) senior faculty involvement and support, (2) utilization of a more holistic approach to graduate admissions, (3) securing financial support for at least one-year of bridging experience that allows students to resolve areas of under preparation in their undergraduate education. (4) providing multiple mentoring resources, (5) being flexible in coursework, (6) sustained progress monitoring and induction into the graduate culture, and (7) a strong research match.  

\section{Summary and outcomes}
\label{sec:outcomes}
  \begin{figure}[ht]
	\begin{center}
			 \epsfig{file=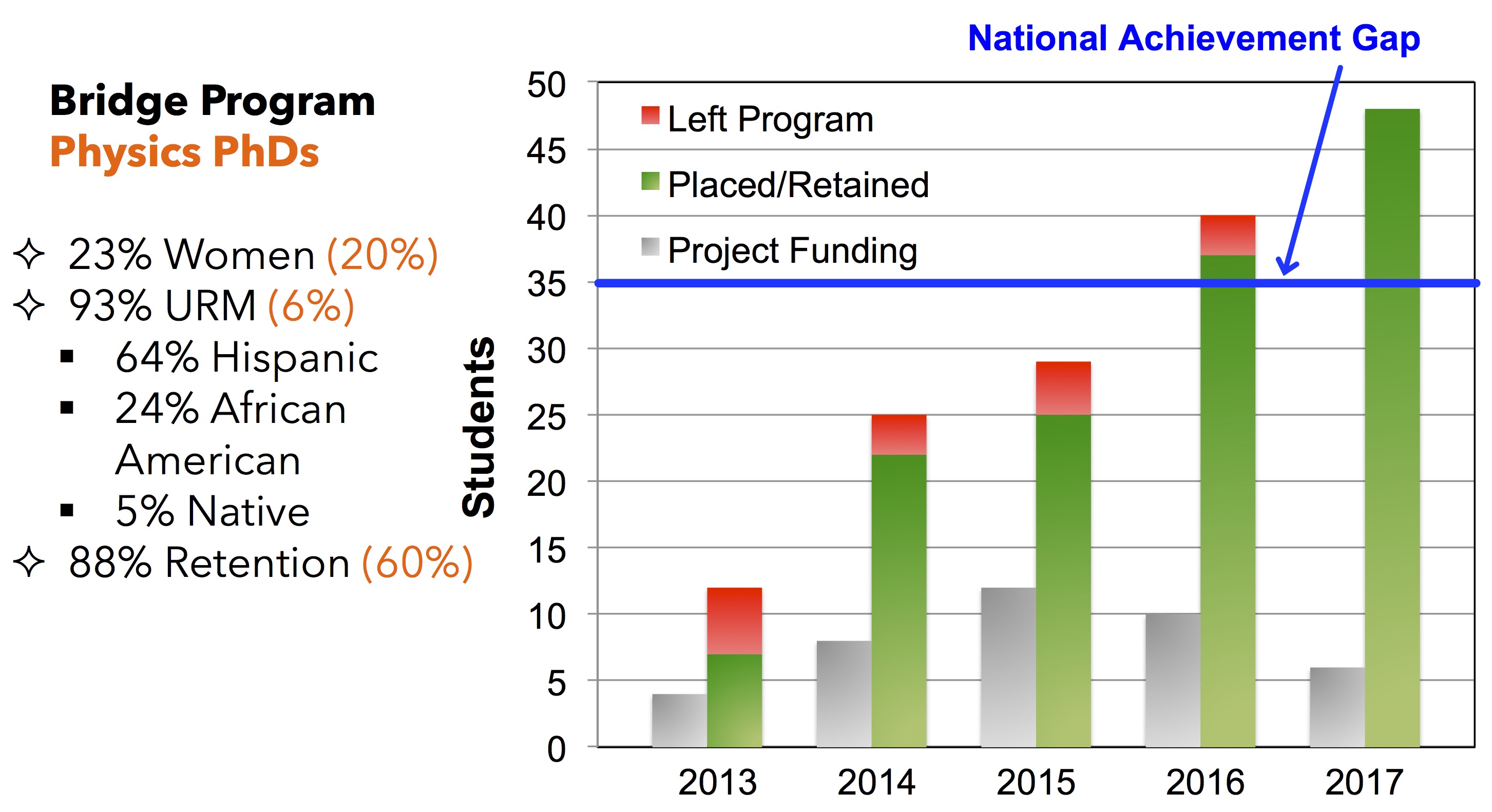,width=0.85\columnwidth}
			\caption{(Color online) Outcomes of APS Bridge Program. The figure shows the number of students placed and retained in Bridge or graduate programs.}
  	\vspace{-.2cm}
	\label{fig:BP_Outcomes}
	\end{center}
\end{figure}

The APS Bridge Program outcomes include 23\% female, and 93\% URM students placed in Bridge or graduate programs. Details of the programs achievements are presented in Figure~\ref{fig:BP_Outcomes}. These are all U.S. students that would not or did not achieve acceptance into a physics graduate program. African-, Hispanic-, and Native-Americans compose 24\%, 64\%, and 5\% respectively of the  93\% URM participation. This level of URM student representation is far higher than the present national averages of female and URMs in U.S. physics graduate departments. A highlight of the APS Bridge Program has been the remarkable retention of  88\% of the placed students, which is substantially higher than the 60\% national average. Retention is defined as students that enter a PhD program in physics and earn their PhD, for the APS Bridge Program it is defined as students making progress toward the PhD. The Fisk-Vanderbilt bridge program has achieved 80\% retention for the students that enter their program, while the University of Michigan's Imes-Moore bridge program in applied physics has surpassed both programs with a 90\% retention rate. Fisk University has established itself as the producer of MS degrees in physics earned by black students, and Vanderbilt is well on its way to becoming a leading producer of doctoral degrees in astronomy, physics, and materials science earned by URM students~\cite{FV}. 

The percentage of bachelor's degrees earned by Hispanic-Americans has grown in correlation with the population increase, in contrast to the persistent decline in the number of bachelor's degrees earned by African Americans over the last 20 years. These students serve as a pipeline for entering into graduate programs and ultimately into research, teaching, industry, or policy positions later on. Bridge Programs are a novel approach at a providing a framework for students earning a BS degrees with a desire to pursue graduate  studies in physics to do so. 
The present achievement clearly demonstrate that Bridge Programs and their practices can be instrumental as an approach to increase diversity in student enrollment, retention, and ultimately the entire physics community. Since establishing its first Bridge Sites and actively recruiting students, the APS Bridge Program has placed over 150 students into Bridge or graduate programs in physics, well beyond the level of the project funding. 

\section{Acknowledgements}
This material is based upon work supported by the National Science Foundation under Grant No. 1143070. Any opinions, findings, and conclusions or recommendations expressed in this material are those of the author(s) and do not necessarily reflect the views of the National Science Foundation.

\def\Discussion{
\setlength{\parskip}{0.3cm}\setlength{\parindent}{0.0cm}
     \bigskip\bigskip      {\Large {\bf Discussion}} \bigskip}
\def\speaker#1{{\bf #1:}\ }
\def\endDiscussion{}

%\Discussion
%\speaker{D. Giovanni (University of Seville)}  My analysis indicates that the recovery of the two gentlemen is due simply to their embrace of the masculine principle and has nothing to do with magnetism at all.  Could you comment on  this?

%\endDiscussion
 
\end{document}